\documentclass[10pt,letterpaper]{article}\usepackage[]{graphicx}\usepackage[]{color}
\pdfoutput=1

\makeatletter
\def\maxwidth{ %
  \ifdim\Gin@nat@width>\linewidth
    \linewidth
  \else
    \Gin@nat@width
  \fi
}
\makeatother

\definecolor{fgcolor}{rgb}{0.345, 0.345, 0.345}

\usepackage{framed}
\makeatletter
 {\par\unskip\endMakeFramed%
 \at@end@of@kframe}
\makeatother

\definecolor{shadecolor}{rgb}{.97, .97, .97}
\definecolor{messagecolor}{rgb}{0, 0, 0}
\definecolor{warningcolor}{rgb}{1, 0, 1}
\definecolor{errorcolor}{rgb}{1, 0, 0}
\newenvironment{knitrout}{}{} 

\usepackage{alltt}

\usepackage{cogsci}
\usepackage{pslatex}
\usepackage{apacite}
\usepackage{amsmath}
\usepackage{amsfonts}
\usepackage{gb4e}
\usepackage{tabulary}

\title{The statistical significance filter leads to 
overconfident expectations of replicability}

 \author{{\large \bf Shravan Vasishth (vasishth@uni-potsdam.de)} \\
 Department of Linguistics, University of Potsdam, Potsdam 14476, Germany.\\
  \AND {\large \bf Andrew Gelman (gelman@stat.columbia.edu)} \\
 Department of Statistics, Columbia University, New York, NY 10027, USA.\\  
  }
\IfFileExists{upquote.sty}{\usepackage{upquote}}{}
\IfFileExists{upquote.sty}{\usepackage{upquote}}{}
\begin{document}






\maketitle

\begin{abstract}
We show that publishing results using the statistical significance filter---publishing only when the p-value is less than 0.05---leads to a vicious cycle of overoptimistic expectation of the replicability of results.
First, we show analytically that when true statistical power is relatively low, computing power based on statistically significant results will lead to overestimates of power. Then, we present a case study using 10 experimental comparisons drawn from a recently published meta-analysis in psycholinguistics (J\"ager et al., 2017). We show that the statistically significant results yield an illusion of replicability.  This illusion holds even if the researcher doesn't conduct any formal power analysis but just uses statistical significance to informally assess robustness (i.e., replicability) of results.

\end{abstract}

\begin{quote}
```\dots in [an]\dots academic environment that only publishes positive findings and rewards publication, an efficient way to succeed is to conduct low power studies. Why? Such studies are cheap and can be farmed for significant results, especially when hypotheses only predict differences from the null, rather than precise quantitative differences and trends.''
\cite[p.\ 5]{NaturalSelection}
\end{quote}

\section{Introduction}

The statistical significance filter tells us that significant results---those findings in which the p-value is less than $0.05$---are positively biased.  The statistically significant estimate is, by definition, more than $t$ standard errors away from zero, where $t$ is some critical value determined by a statistical test (such as the t-test) and the pre-specified Type I error (the probability, under repeated sampling, of incorrectly rejecting the null hypothesis). 

Statistical power is the probability, under repeated sampling, of correctly rejecting the null hypothesis 
assuming that the parameter of interest has some true point value $\mu$.\footnote{In order to compute power, we need to have an estimate of the true effect, the sample size, and an estimate of the standard deviation.}
It is well-known that when statistical power is low, the effect (the sample mean) will tend to be exaggerated. These are referred to as Type M errors by \citeA{gelmancarlin} (also see \citeNP{gelman2000type}). This exaggeration of effects has been noticed in previous work \cite{hedges1984estimation,lane1978estimating}, and 
most recently in neuroscience and epidemiology, where
\citeA{powerfailure} refer to the exaggeration of effects in neuroscience as the ``winner's curse'' and ``the vibration of effects.'' In related work, \citeA{ioannidis2008most} discusses this exaggeration of effects in epidemiological studies in terms of the vibration ratio: the ratio of largest to smallest observed effects.

These overestimates get published and fill the literature. Now consider what happens when researchers design a new study. They read the literature and see all these big effects, then plan their next study. They do a power calculation based on these big effects and get an exaggerated estimate of power, and can easily convince themselves that they have a high powered study. Alternatively---and this is probably the more common route in many fields, such as psychology---they don't do a formal power analysis, but just rely on the informal observation that most of the previously published results had a significant effect and so the effect must be present.

A related observation about overestimation comes from the replication attempts reported by the \citeA{open2015estimating}. 
The authors report that the magnitude of the published p-values from the original studies were predictive of replication success. As they put it (p.\ 943): ``\dots correlational evidence is consistent with the conclusion that variation in the strength of initial evidence (such as original P value) was \dots predictive of replication success \dots''
From this, researchers might erroneously conclude that lower p-values are
generally more predictive of replication success. In other words, an erroneous conclusion would be that a lower p-value suggests a higher probability that the effect can be detected in future repeated studies.

We show that if statistical significance is used as a filter for publishing a result, and the observed effect (or p-value) is used to determine replicability, this will lead the researcher to overestimate replicability.  We demonstrate this point analytically, and then present a case study involving 10 reading studies in psycholinguistics that illustrates this illusion.

\section{The relationship between p-values and estimated power}

Assume for simplicity the case that we carry out a one-sided statistical test where the null hypothesis is that the null hypothesis mean is $\mu_0=0$ and the alternative is that $\mu>0$.\footnote{The presentation below generalizes to the two-sided test.}
Given some continuous data $x_1,\dots,x_n$, we can compute the t-statistic and derive the p-value from it. For a large sample size $n$, a normal approximation allows us to use the z-statistic, $Z=\frac{\bar X-\mu_0}{ \sigma_X/\sqrt{n}}$, to compute the p-value. Here, $\bar{X}$ is the mean,  $\sigma_X$ the standard deviation, and $n$ the sample size. 

The p-value is the probability of observing the z-statistic or a value more extreme assuming that the null hypothesis is true.
The p-value is a random variable $P$ with the probability density function \cite{hung1997behavior}: 

\begin{equation}
g_\delta(p) = \frac{\phi(Z_p - \delta)}{\phi(Z_p)},\quad 0<p<1
\end{equation}

\noindent
where 
\begin{itemize}
\item
$\phi(\cdot)$ is the pdf of the standard normal distribution, Normal(0,1).
\item
$Z_p$, a random variable, is the (1-p)th percentile of the standard normal distribution. 
\item
$\delta=\frac{\mu-\mu_0}{\sigma_X/\sqrt{n}}$ is 
the true point value expressed as a z-score. Here, $\mu$ is the true (unknown) point value of the parameter of interest.
\end{itemize}

\citeA{hung1997behavior} further observe that the cumulative distribution function (cdf) of $P$ is:

\begin{equation}
G_\delta(p) = \int_0^p g_\delta(x)\, dx = 1- \Phi(Z_p-\delta),\quad 0<p<1
\end{equation}

\noindent
where $\Phi(\cdot)$ is the cdf of the standard normal.

Once we have observed a particular z-statistic $z_p$, 
the cdf $G_\delta(p)$ allows us to estimate power based on the z-statistic \cite{hoenigheisey}.
To estimate the p-value given that the null hypothesis is true, let the true value be $\mu=0$. It follows that $\delta=0$. Then:

\begin{equation}
p = 1-\Phi(z_p)
\end{equation}

To estimate power from the observed $z_p$, set $\delta$ to be the observed statistic $z_p$, and let the critical z-score be $z_\alpha$, where $\alpha$ is the Type I error (typically $0.05$).
The power is therefore:

\begin{equation} \label{powerequation}
G_{z_p}(\alpha) = 1- \Phi(z_\alpha - z_p)
\end{equation}

\begin{figure}[!htbp]
\centering
\begin{knitrout}
\definecolor{shadecolor}{rgb}{0.969, 0.969, 0.969}\color{fgcolor}
\includegraphics[width=\maxwidth]{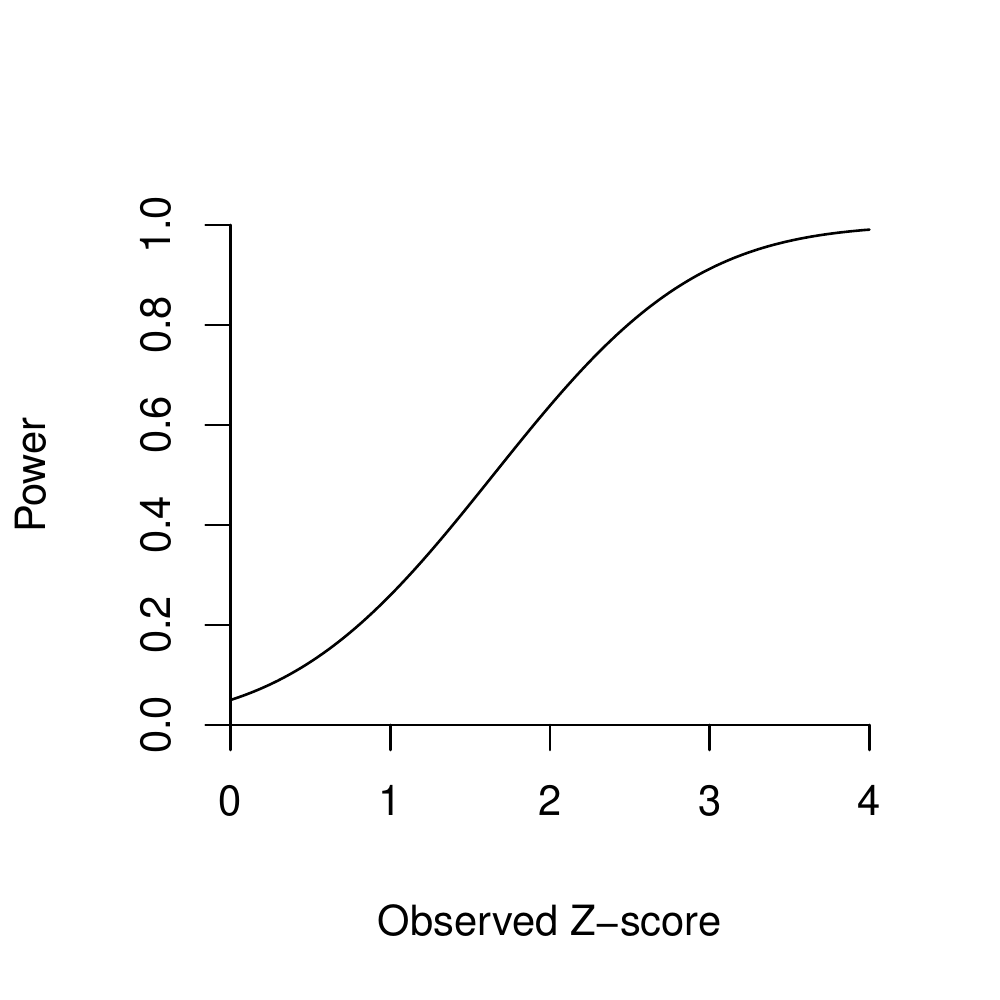} 

\end{knitrout}
\caption{The relationship between power and the observed z-score. The larger z-scores are easier to publish due to the statistical significance filter, and these published studies therefore give a mistaken impression of higher power.}\label{fig:powerz}
\end{figure}

In other words, power estimated from the observed statistic is a monotonically increasing function of the observed z-statistic: the larger the statistic, the higher the power estimate based on this statistic (Figure~\ref{fig:powerz}). 
Together with the common practice that only statistically significant results get published, and especially results with a large z-statistic, this leads to overestimates of power. As mentioned above, one doesn't need to actually estimate power in order to fall prey to the illusion; merely  scanning the statistically significant z-scores gives an impression of consistency and invites the inference that the effect is replicable and robust. The word ``reliable'' is frequently used in psychology, presumably with the meaning that the result is replicable and represents the reality. 

A direct consequence of Equation~\ref{powerequation} is that overestimates of the z-statistic will lead to overestimates of power.
For example, if we have $36$ data points and the true effect is $0.1$ on some scale and standard deviation is $1$, then
statistical power is $15$\%.\footnote{This can be confirmed by running the following command using R \cite{R}: \texttt{power.t.test(delta=0.1,sd=1,n=36,alternative = "one.sided",type="one.sample")}.}



%

If we now re-run the same study, collecting $36$ data points each time, and impose the condition that only statistically significant results with  Type I error $\alpha=0.05$ are published, then only observed z-scores larger than $1.64$ (for  a one-sided test) would be published and the power estimate based on these z-scores must have a lower bound of

\begin{equation}
G_{Z_\alpha}(\alpha) = 1- \Phi(1.64-1.64)=0.5
\end{equation}

\noindent
Thus, in a scenario where the real power is 15\%, and only z-scores greater than or equal to $z_\alpha$ are published, 
power based on the z-score will be overestimated by at least a factor of 0.5/0.15=3.33. Call this ratio the Power Inflation Index (PII). 

Now, lower p-values are widely regarded as more ``reliable'' than p-values near the Type I error probability of $0.05$.\footnote{Treating lower p-values as furnishing more evidence against the null hypothesis reflects a misunderstanding about the meaning of the p-value; given a continuous dependent measure, when the null hypothesis that $\mu=0$ is true, under repeated sampling the p-value has a uniform distribution (see proof in the Appendix). This has the consequence that, when the null is true, a p-value near 0 is no more surprising than a p-value near 0.05.}    
This incorrect belief, widely shared by editors, reviewers, and authors in areas like psychology and linguistics, has the effect that studies with lower p-values are more likely to be reported and published, with the consequence that the PII will tend to be even higher than the lower bound discussed here.

We turn next to a case study involving psycholinguistic data that illustrates the illusion of replicability.

\section{Case study: Interference effects in reading studies}

To illustrate the illusion of replicability, we consider the $10$ experiments that were reviewed in the literature review and meta-analysis presented in \citeA{JaegerEngelmannVasishth2017}.
These were psycholinguistic studies in which the dependent measure was reading time in milliseconds of words. The experimental manipulation involved pairs of sentence types where one type was easier to read than the other; the empirical phenomenon of interest here is interference in working memory. Here, an appropriate statistical test is the two-sided paired t-test (one could do a one-sided t-test, although this is less common in psycholinguistics). 

We had the raw data from these $10$ studies and so were able to carry out the pairwise comparison. As discussed in detail in \citeA{JaegerEngelmannVasishth2017}, theory
predicts an effect with a negative sign. The original results as published were analyzed on the raw milliseconds scale, but here we analyze the data on the log milliseconds scale
because the reading time data were log-normally distributed. 

A summary of the pairwise t-test is shown in Table~\ref{tab:ttests}. From the table, it is clear that the studies consistently found negative values for the coefficient; this consistent result raises our confidence in the reproducibility of the result. A formal power analysis based on these studies, also shown in the last column of the table, 
leads to estimates of power ranging from 17 to 60\%.

\begin{table}[ht]
\centering
\begin{tabular}{rrrrrrrr}
  \hline
 & t & d & n & se & s & pval & power \\ 
  \hline
1 & -1.9 & -0.1 & 40 & 0.0 & 0.2 & 0.1 & 0.3 \\ 
  2 & -3.1 & -0.1 & 32 & 0.0 & 0.1 & 0.0 & 0.6 \\ 
  3 & -1.5 & -0.0 & 32 & 0.0 & 0.2 & 0.2 & 0.2 \\ 
  4 & -2.1 & -0.0 & 32 & 0.0 & 0.1 & 0.0 & 0.3 \\ 
  5 & -1.7 & -0.0 & 32 & 0.0 & 0.1 & 0.1 & 0.2 \\ 
  6 & -2.6 & -0.1 & 28 & 0.0 & 0.2 & 0.0 & 0.4 \\ 
  7 & -1.6 & -0.0 & 60 & 0.0 & 0.2 & 0.1 & 0.2 \\ 
  8 & -3.2 & -0.1 & 44 & 0.0 & 0.2 & 0.0 & 0.6 \\ 
  9 & -1.9 & -0.1 & 60 & 0.0 & 0.2 & 0.1 & 0.3 \\ 
  10 & -2.6 & -0.0 & 114 & 0.0 & 0.2 & 0.0 & 0.5 \\ 
   \hline
\end{tabular}
\caption{Results from the paired t-tests for the 10 experimental comparisons. Shown are the t-score, the effect d in log ms, the sample size n, the standard error se, the standard deviation s, and the p-value. The t-tests were done on the raw data from the original studies (the t-values reported here may deviate slightly from the published t-values). Also shown is the power estimated from each study.} 
\label{tab:ttests}
\end{table}

\subsection{Using a Bayesian random-effects meta-analysis to estimate the power function}

In Table~\ref{tab:ttests}, 
we calculated power based on the individual studies. As discussed above, 
these will tend to be overestimates because there is a preference to publish effects with low p-values. How can we check this for the 10 studies? True power is unknown so we have no basis for comparing the power estimates from individual studies with a true value for power.

One way to arrive at a conservative estimate of the true power given these 10 studies
is to carry out a Bayesian random-effects meta-analysis \cite{Gelman14}. This hierarchical modelling approach allows us to determine the posterior distribution of the effect, which can then be used for computing an estimate of power. As discussed in \citeA{powerfailure}, using estimates from a meta-analysis yields a more conservative estimate of power. In the random-effects meta-analysis, this conservativity arises due to the shrinkage property of hierarchical models: Larger sample studies receive a greater weighting in determining the posterior than smaller sample studies.
Note, however, that even here the power may be an overestimate due to the fact that the studies that go into the meta-analysis are likely to have publication bias. But as we show below, the estimates  of power from individual studies tend to be ever larger.

The random-effects meta-analysis model was set up as follows. Let $y_i$ be the effect size in log milliseconds in the $i$-th study, where $i$ ranges from $1$ to $n$.  Let $\mu$ be the true (unknown) effect in log ms, to be estimated by the model, and $\mu_i$ the true (unknown) effect in each study.
Let $\sigma_{i}$~log ms be the true standard deviation of the sampling distribution; each $\sigma_i$ is estimated from the sample standard error from study $i$. The standard deviation parameter $\tau$ represents between-study variability.

Then, our model for $n$ studies is as follows. The model assumes the i-th data point (the effect observed on the log ms scale) $y_i$ is generated from a normal distribution with mean $\mu_i$ and some standard error $\sigma$, estimated from the sample's standard error. Each of the true underlying means $\mu_i$ are assumed
to be generated from a normal distribution with true mean $\mu$ and between-study standard deviation $\tau$. We assign Cauchy(0,2.5) priors to the parameters $\mu$ and $\mu_i$,  and  a truncated Cauchy(0,2.5) prior for the between-study standard deviation
$\tau$, truncated so that $\tau$ is greater than $0$. 
The model can be stated mathematically as follows:

\begin{equation}
\begin{split}
\hbox{Likelihoods:} & \\
y_i \mid \mu_i, \sigma_i^2 \sim & Normal(\mu_i, \sigma_i^2) \quad i=1,\dots, n\\
\mu_i\mid \theta,\tau^2 \sim & Normal(\mu, \tau^2), \\
\hbox{Priors:} & \\
\mu \sim & Cauchy(0,2.5),\\
\mu_i \sim & Cauchy(0,2.5),\\
\tau \sim & Cauchy(0,2.5), \tau>0 \\
\end{split}
\end{equation}

\begin{figure}[!htbp]
\centering
\begin{knitrout}
\definecolor{shadecolor}{rgb}{0.969, 0.969, 0.969}\color{fgcolor}
\includegraphics[width=\maxwidth]{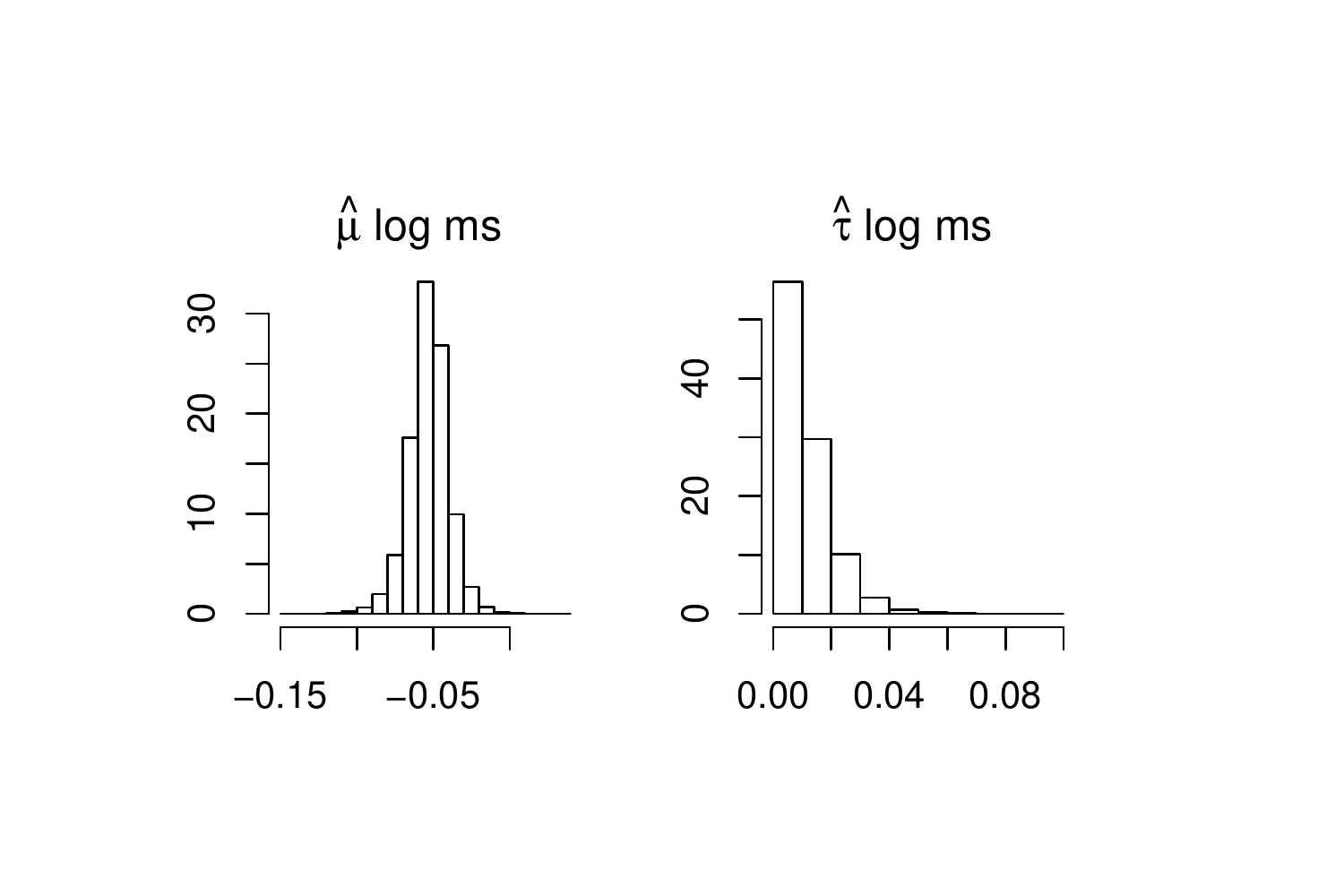} 

\end{knitrout}
\caption{Posterior distributions of the estimated effect ($\hat\mu$), and the standard deviation of estimate of the between-study variability ($\hat\tau$) in the random-effects meta-analysis.}\label{fig:rema}
\end{figure}

We fit the model using Stan 2.14.2 \cite{stan-manual:2016}, running four chains with 4000 iterations (half of which were warm-ups). Convergence was successful, as diagnosed using the $\hat R$ diagnostic \cite{Gelman14}.  The posterior distributions of $\hat\mu$ and of the between-study standard deviation $\hat\tau$ are shown in Figure~\ref{fig:rema}.
The posterior mean of the effect is -0.05 log ms, with 95\% credible interval [-0.08,-0.03].
Next, we use this estimate of the posterior distribution to compute a power distribution.

\subsubsection{Computing the power distribution using the posterior distribution of the effect}

An analysis of reading studies, including the ones considered here, showed that the precisions (the inverse of the variance) in reading time studies 
have mean values $16.3$ and standard deviation $7.07$ (the unit for precision is  1/log ms$^2$). Since precision can be modelled as a Gamma distribution, 
we assumed that precisions  are distributed as $Gamma(\alpha=5.3,\beta=0.3)$. These parameters of the Gamma distribution were computed by taking the mean $\bar{x}$ and standard deviation $s$ of the precisions, and then deriving the parameters of the Gamma distribution by solving for $\alpha$ and $\beta$. 
We use the fact that 
for a random variable generated from a Gamma distribution with parameters $\alpha$ and $\beta$, the expectation $\mu$ and variance $\sigma^2$ are:

\begin{equation}
E(X) = \frac{\alpha}{\beta}=\mu \quad \hbox { and } Var(X)= \frac{\alpha}{\beta^2}= \sigma^2
\end{equation}






Having obtained the estimate of the effect (through the meta-analysis) and the distribution of the precisions, we used these estimates to carry out $100,000$ Monte Carlo simulations to derive a power distribution for different sample sizes ($n=20,\dots,50$) in the following manner. For each sample size, we repeatedly computed power after obtaining: 

\begin{itemize}
\item one sample for the effect by sampling from the distribution $Normal(-0.05,0.01)$; this is the posterior distribution of the effect derived from the random-effects meta-analysis; 
\item one sample  for the precision by sampling from the $Gamma(5.3,0.3)$, and then converting this to a standard deviation.
\end{itemize}

\noindent
Such a Monte Carlo sampling procedure gives a probability distribution of power values and allows us to quantify our uncertainty about the estimated power by taking all sources of uncertainty into account---the uncertainty regarding the effect, and the uncertainty regarding the standard deviation. 

\begin{figure}[!hbtp]
\centering

\begin{knitrout}
\definecolor{shadecolor}{rgb}{0.969, 0.969, 0.969}\color{fgcolor}
\includegraphics[width=\maxwidth]{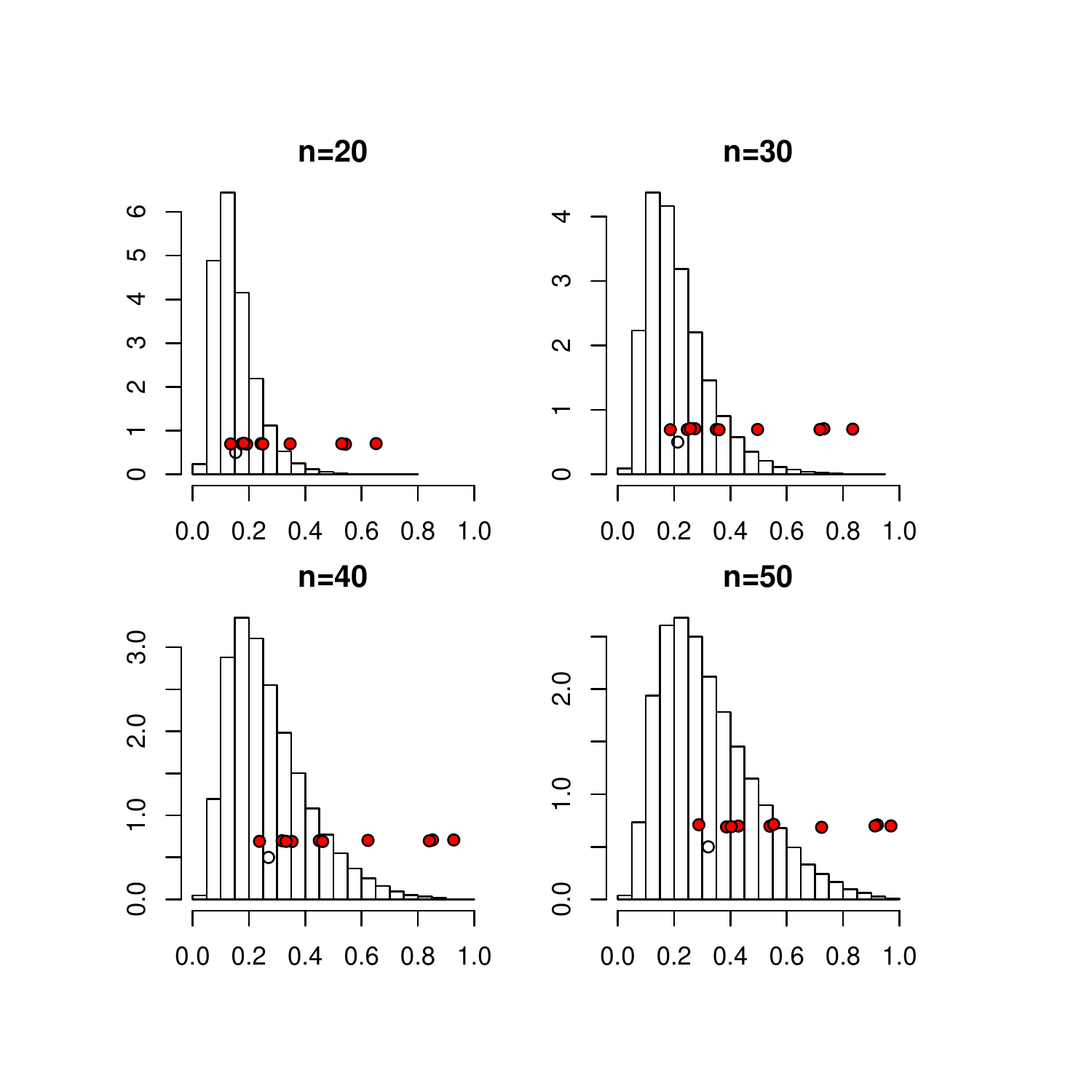} 

\end{knitrout}
\caption{Power distributions for different sample sizes (log reading times). The histogram shows the power distribution (generated through Monte Carlo sampling; see text for details). The red dots show power estimates from the 10 individual experimental comparisons considered in this case study. The white dot shows the mean of each power distribution.}\label{fig:powerdistrnlog}
\end{figure}

Figure~\ref{fig:powerdistrnlog} shows the resulting power distributions for power given different sample sizes.
These power distributions are of course only estimates, not the true power; and as \citeA{powerfailure} point out, are probably slight overestimates if the studies themselves have publication bias.

The power distributions illustrate two important points. First, the 
range of most likely power values is remarkably low for typical sample sizes used in psycholinguistic reading experiments relating to interference effects (see Table~\ref{tab:ttests}). 
As an aside, we note that our estimates are similar to those from a recent review of 44 meta-analyses of research in social and behavioural sciences published between 1960-2011; they report a mean power of 0.24 with most studies suggesting power to be below 0.4 \cite[p.\ 6, Fig.\ 1]{NaturalSelection}. 
The second observation is that  the power values computed from individual studies (the red dots) tend to be overestimates relative to the mean of each power distribution shown. The power from each study tends to be higher than the mean of each power distribution. Of course, if the statistical power of the original studies were very high (approximately 80\% or higher), then the overestimation problem would disappear or at least be negligible.

We can quantify the overestimation of power by computing the Power Inflation Index: the ratio of the power computed from individual studies to the power distribution computed using Monte Carlo simulations. If power is overestimated, then the distribution of the PII will be such that the mean ratio will be greater than 1. 
These distributions of PIIs are computed for a typical  sample size used in psycholinguistic studies (n=20, 30, 40) in Table~\ref{tab:PII}. Here, we can see that the PII can be as high as 12. 

\begin{table}[!htbp]
\centering
\begin{tabular}{rrrrrrr}
  \hline
      & \multicolumn{2}{c}{n=20} & \multicolumn{2}{c}{n=30} & \multicolumn{2}{c}{n=40}\\
Study & 2.5\% & 97.5\% & 2.5\% & 97.5\% & 2.5\% & 97.5\% \\ 
  \hline
1 & 1.37 & 4.64 & 0.98 & 3.95 & 0.76 & 3.47 \\ 
  2 & 3.67 & 12.45 & 2.63 & 10.61 & 2.04 & 9.30 \\ 
  3 & 1.08 & 3.67 & 0.78 & 3.13 & 0.60 & 2.74 \\ 
  4 & 1.95 & 6.61 & 1.40 & 5.64 & 1.08 & 4.94 \\ 
  5 & 1.41 & 4.76 & 1.01 & 4.06 & 0.78 & 3.56 \\ 
  6 & 3.06 & 10.37 & 2.19 & 8.83 & 1.70 & 7.75 \\ 
  7 & 0.76 & 2.59 & 0.55 & 2.20 & 0.42 & 1.93 \\ 
  8 & 2.99 & 10.12 & 2.14 & 8.62 & 1.65 & 7.56 \\ 
  9 & 0.98 & 3.34 & 0.71 & 2.84 & 0.55 & 2.49 \\ 
  10 & 1.02 & 3.47 & 0.73 & 2.96 & 0.57 & 2.59 \\ 
   \hline
\end{tabular}
\caption{The 95\% credible intervals of the Power Inflation Index for each of the 10 experimental comparisons, for different sample sizes. The Power Inflation Index can be as large as 12.}\label{tab:PII}
\end{table}

\section{Discussion}

We have shown that if statistical significance is used to decide whether to publish a result, overestimates of the effect will tend to be published, leading to an over-enthusiastic belief in the replicability of the effect.

Recently, the replication project reported by \citeA{open2015estimating} showed that only 47\% of the studies they investigated could be replicated. One factor causing these failures to replicate could have been low power in the original studies. Even before the replication project, \citeA{cohen1962statistical,powerbookcohen} and others have repeatedly warned against running low-powered studies. Despite these injunctions, many researchers do not believe that there is a problem of low power.  
For example, \citeA{Gilbert1037} contested the 47\% replication rate and argued that the replication rate may be much higher, perhaps even ``statistically indistinguishable from 100\%.'' The objections of \citeA{Gilbert1037} were largely based on arguments about the lack of fidelity to the original design, but it is possible that, in addition to concerns about fidelity,
Gilbert et al.\ are, like many researchers, generally overconfident about the replicability and robustness of their results. This overconfidence is also evident in reading research in psycholinguistics, where it is routine to run experiments with sample sizes ranging from 20 to 40 participants. Recent work has argued that sample sizes of 20-40 partipants may be too low for reading studies on interference \cite{JaegerEngelmannVasishth2017}. We are hopeful that future work will take this finding into account when planning studies. 

Currently, the replication problems in psycholinguistics are serious.
For example, in recent work \cite{MertzenEtAl2017CUNY} we carried out six replication attempts of two eyetracking experiments published in the \textit{Journal of Memory and Language}.  
 We were unable to replicate any of the claims in the paper. 
There is thus an urgent need to attempt to replicate published results, and not just in psycholinguistics. For example,  \citeA{makel2012replications} present a quantitative analysis of the low rate of successful replications in psychology (1\%).
Other fields are also affected. For example, \citeA{powerfailure} have shown that in neuroscience studies, power may also be quite low, ranging from 8 to 31\%. \citeA{NaturalSelection} have shown through a 50-year meta-analysis in  behavioural science that power has not improved (mean power: 24\%). In biomedical sciences, approximately 50\% of studies have power in the 0-10\%\footnote{Note that this range is an error; power cannot be less than 5\% if Type I error is set at 5\%.} or 11-20\% range \cite{PowerBioMed}. 

Despite these indications, many researchers remain overconfident about the robustness of their results. This overconfidence is in part due to the statistical significance filter. 


\section{Concluding remarks}

We have shown that the statistical significance filter directly leads to over-optimistic expectations of replicability of published research. Even if the researcher doesn't conduct any formal power analyses, they can fall prey to this illusion because of the informal assessment of replicability afforded by the statistical significance filter. We illustrated the illusion of replicability through a case-study
involving 10 published experimental comparisons.

Many psychology journals are beginning to require that power analyses be included in submitted manuscripts. But our results, echoing those of others who have studied this problem, suggest that such analyses, which invariably are based on previously published work, will tend to provide overestimates of power.

To resolve or at least reduce this problem, we offer two pieces of advice.  First, we recommend entirely abandoning the concept of power, which is based on the idea that ``$p < .05$'' is a win, an attitude that fails miserably when effect sizes are small and measurements are noisy.  Second, when performing design analysis, consider possible effect sizes based on subject-matter understanding; see \citeA{gelmancarlin} for further discussion of this point.  It can make sense to consider a range of reasonable effect sizes.

\section*{Appendix}
Here, we review the well-known proof  that for a point null hypothesis and a continuous dependent variable, the distribution of the p-value under the null is $Uniform(0,1)$.

When a random variable $Z$ comes from a $Uniform(0,1)$ distribution, then the probability that $Z$ is less than (or equal to) some value $z$ is exactly $z$: $P(Z\leq z)=z$.

The p-value is a random variable, call it $Z$. The p-value is computed by calculating the probability of seeing a t-statistic or something more extreme under the null hypothesis. The t-statistic comes from a random variable $T$ that is a transformation of the random variable $\bar{X}$: $T=(\bar{X}-\mu)/(\sigma/\sqrt{n})$. This random variable T has a CDF $F$.

We can establish that
if a random variable $Z=F(T)$, then $Z \sim Uniform(0,1)$, i.e., that
 the p-value's distribution under the null hypothesis is $Uniform(0,1)$.
This is proved next.

Let $Z=F(T)$. Then:
$P(Z\leq z) = P(F(T)\leq z) = P(F^{-1} F(T) \leq F^{-1}(z) )
= P(T \leq F^{-1} (z) )
= F(F^{-1}(z))= z$.

Since $P(Z\leq z)=z$, Z is uniformly distributed, that is, $Uniform(0,1)$.



\section*{Acknowledgements}

We thank Lena J\"ager,
Reinhold Kliegl, Christian Robert, and Bruno Nicenboim for helpful discussions.
For partial support of this research, 
we thank 
the Volkswagen Foundation through grant 89 953, and 
the U.S.\ Office of Naval Research  through grant N00014-15-1-2541.

\bibliographystyle{apacite}

\setlength{\bibleftmargin}{.125in}
\setlength{\bibindent}{-\bibleftmargin}

\bibliography{/Users/shravanvasishth/Dropbox/Bibliography/bibcleaned}

\end{document}